\documentclass[aps,prb,twocolumn,floatfix,superscriptaddress,longbibliography]{revtex4-2}
\usepackage[utf8]{inputenc}
\usepackage[english]{babel}
\usepackage{mathtools}
\usepackage{newtxtext}  
\usepackage{amsmath}
\usepackage{amsfonts}
\usepackage[varbb,varg]{newtxmath} 
\usepackage{bm,dsfont,eucal}
\AtBeginDocument{\renewcommand{\vec}[1]{\bm{#1}}}
\newcommand{\pdagger}{\phantom{\dagger}}

\usepackage[dvipsnames,svgnames,x11names,hyperref]{xcolor}
\usepackage{hyperref}
\hypersetup{ 
    colorlinks=true, linkcolor=NavyBlue, urlcolor=NavyBlue, citecolor=NavyBlue
}

\usepackage{physics,braket}
\usepackage[capitalise]{cleveref}

\begin{document}

\title{Quasi-boson approximation yields accurate correlation energy in the 2D electron gas} 

\author{Tobias M.~R.~Wolf}
\affiliation{Department of Physics, University of Texas at Austin, Austin, Texas 78712, USA}
\author{Chunli Huang}
\affiliation{Department of Physics and Astronomy, University of Kentucky, Lexington, Kentucky 40506-0055, USA}

\date{\today} 

\begin{abstract}
We report the successful adaptation of the quasi-boson approximation, a technique traditionally employed in nuclear physics, to the analysis of the two-dimensional electron gas.  We show that the correlation energy estimated from this approximation agrees closely with the results obtained from quantum Monte Carlo simulations. Our methodology comprehensively incorporates the exchange self-energy, direct scattering, and exchange scattering for a particle-hole pair excited out of the mean-field groundstate within the equation-of-motion framework. The linearization of the equation of motion leads to a generalized-random-phase-approximation (gRPA) eigenvalue equation whose spectrum indicates that the plasmon dispersion remains unaffected by exchange effects, while the particle-hole continuum experiences a marked upward shift due to the exchange self-energy.  Notably, the plasmon mode retains its collective nature within the particle-hole continuum, up to moderately short wavelength ($q\sim 0.3 k_F$ at metallic density $r_s=4$). 
Using the gRPA excitation spectrum, we calculate the zero-point energy of the quasi-boson Hamiltonian, thereby approximating the correlation energy of the original Hamiltonian.
This research highlights the potential and effectiveness of applying the quasi-boson approximation to the gRPA spectrum, a fundamental technique in nuclear physics, to extended condensed matter systems.
\end{abstract}

\maketitle

\section{Introduction} \label{sec:intro}

The correlation energy, defined as the difference between the exact groundstate energy and the self-consistent Hartree-Fock energy, plays an important role in shaping the phase diagrams of various quantum many-body systems within nuclear and condensed matter physics \cite{thouless2014quantum,rowe2010nuclear}. The success in estimating the correlation energy in nuclear matter for a given nucleon-nucleon interaction profile is largely attributed to the quasi-boson approximation, a method notable for its adaptability and precision. Recent advances have seen this approximation evolve in sophistication and application, as discussed in Ref.~\cite{schuck2021equation}.
Despite its success in nuclear matter, the quasi-boson approximation has not been extensively applied to metals. This study seeks to bridge this gap by applying the quasi-boson approximation to the two-dimensional electron gas (2DEG) system.

The electron gas model has a long history \cite{giuliani2005quantum,pines1990theory,herring1966magnetism,gell1957correlation,nozieres1958correlation,bohm1953collective,rajagopal1977correlations}, with the first estimation of correlation energy traced back to the random-phase approximation (RPA) introduced by Bohm and Pines \cite{bohm1953collective}. 
The RPA only retains terms in the diagrammatic perturbation series where different Fourier components of the Coulomb interaction, $V_q$, are uncoupled. The RPA is an excellent approximation for describing long-distance ($q\rightarrow0$) collective phenomena such as screening since it contains the most diverging geometric series in perturbation theory. However, its efficacy diminishes for short-distance phenomena, where exchange scattering and the particle-particle ladder processes become important.
To overcome the limitations of the RPA, an approach often employed introduces the so-called local-field factors into the particle-hole response function \cite{hubbard1958localfield,singwi1968stls, jonson1976electron,schulze2000two}. This method incorporates short-distance correlations while preserving the original RPA structure, which makes it eminently practical. The first inception of this idea dates back to Hubbard \cite{hubbard1958localfield} who observed that the exchange contribution of a diagram in the perturbation series will tend to cancel one-half of the direct contribution. He used this insight to approximately evaluate an infinite series of bubble-exchange diagrams and compute the correlation energy. It was later refined into the so-called STLS method \cite{singwi1968stls} that uses a self-consistent semiclassical approximation for the local-field factors.  Currently, the quantum Monte Carlo \cite{schulze2000two,tanatar1989ground} calculation is recognized as providing the most accurate estimation of correlation energy, setting a benchmark for comparison in this field.

\begin{figure}
\centering
\includegraphics[width=1.0\columnwidth]{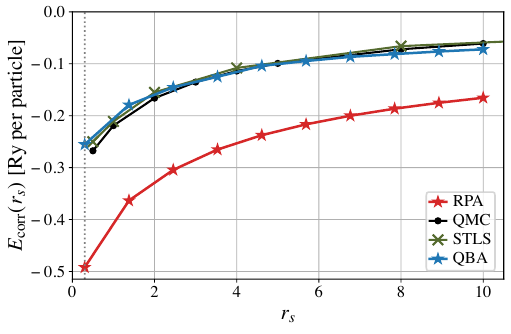}
\caption{
Correlation energy $v.s.~$ electron gas parameter $r_s$ in the 2DEG obtained in the quasi-boson approximation (QBA) for the generalized RPA spectrum. The quasi-boson approximation yields results in good agreement with the Quantum Monte Carlo (QMC) and the local-field method (STLS) over a wide range of $r_s$. The data for QMC and STLS are obtained from Refs.~\cite{schulze2000two,jonson1976electron}. For QBA, we used a Yukawa screening wavevector of $\kappa=0.17 dk$, where $dk$ is the discretization.
}
\label{fig1}
\end{figure}

In this study, we adopt an equation of motion approach to numerically sum over the complete geometric series of bubble-exchange diagrams, thereby obtaining the particle-hole excitation spectrum. Subsequently, we leverage this spectrum to compute the correlation energy. This computation is based on the assumption that the commutator of the particle-hole annihilation operator ($b_{ni}$) and the particle-hole creation operator ($b_{mj}^\dagger$) can be replaced by its Hartree-Fock (HF) expectation value, i.e.,
\begin{align} \label{eq:QBA}
    [b_{ni},b_{mj}^\dagger] \approx  \langle \textrm{HF}|[b_{ni},b_{mj}^\dagger]  | \textrm{HF}\rangle =\delta_{mn}\delta_{ij},
\end{align}
where $m,n$ ($i,j$) label particle (hole) states and $\ket{\textrm{HF}}$ is the Hartree-Fock ground state. This is known as the quasi-boson approximation (QBA) because the above equation would have been exact if $b_{ni}$ and $b_{mj}^\dagger$ truly were bosonic operators. Our results demonstrate that the correlation energy calculated within the QBA is in good agreement with quantum Monte Carlo (QMC) results, as shown in \cref{fig1}. This intriguing finding suggests that while the exact 2DEG groundstate $|\Psi\rangle$ is surely orthogonal to the Hartree-Fock groundstate, $\langle \Psi|\text{HF}\rangle \approx 0$, the expectation value of the above commutator is well approximated by its HF value: 
\begin{align}
    \langle \Psi |[b_{ni},b_{mj}^\dagger] |\Psi \rangle  \approx  \langle \textrm{HF}[b_{ni},b_{mj}^\dagger]|\textrm{HF}\rangle.
\end{align}
In terms of the many-body wavefunction, the QBA introduces \textit{zero-point} quantum fluctuation into the HF groundstate 
\footnote{Within the nuclear physics community, the QBA is often just called RPA, even though it actually contains the exchange term as well. In the language of Ref.~\cite{schuck2021equation}, our work employs standard RPA, as opposed to self-consistent RPA.}
While the QBA is a widely-used method in nuclear many-body physics \cite{peter1980nuclear,schuck2021equation,delion2016}, it remains largely uncharted in condensed matter physics, likely due to its computational complexities.

This paper is structured as follows: Section \cref{sec:method} revisits the application of the equation of motion to derive the generalized RPA equation. We then numerically solve this equation to obtain the excitation eigenspectrum and response functions of the electron gas across varying density parameters $r_s$. We discuss the impact of exchange self-energy and exchange scattering on the plasmon dispersion and particle-hole continuum.
In \cref{sec:corr_energy}, we employ a quasi-boson approximation to obtain the 2DEG correlation energy as ground state energy of a bosonic Hamiltonian for residual interactions. Making use of the gRPA eigenspectrum then allows us to evaluate the quasi-boson correlation energy at different $r_s$.
We conclude with a discussion and outlook in \cref{sec:conclusion}.
 
\section{Equation of motion approach}\label{sec:method}

In this section, we use the equation of motion approach to derive the generalized random phase approximation (gRPA) eigenvalue equation and compute excitation spectrum associated with the creation of a single particle-hole pair in 2DEG. To remind the reader, the 2DEG Hamiltonian in momentum space is
\begin{align} \label{eq:H_toe}
    H=\sum_{\vec{k}\sigma} \epsilon_{\vec{k}} \, c^{\dagger}_{\vec{k}\sigma}c^{\pdagger}_{\vec{k}\sigma}
    +\frac{1}{2}\sum_{\vec{q}\neq0}V_{\vec{q}} \, :\rho_{\vec{q}} \rho_{-\vec{q}}:, 
\end{align}
where $\epsilon_{\vec{k}}=\hbar^2\vec{k}^2/2m$ is the dispersion, $c_{\vec{k}\sigma}^{(\dagger)}$ destroys (creates) an electron with spin $\sigma$ and momentum $\vec{k}$, $\rho_{\vec{q}} = \sum_{\vec{k}\sigma} c^{\dagger}_{\vec{k}+\vec{q}\sigma}c^{\pdagger}_{\vec{k}\sigma}$ is the density operator, $V_{\vec{q}}=2\pi e^2/A\sqrt{\vec{q}^2+\kappa^2}$ is the Fourier component of (screened) Coulomb potential, ${:\cdot:}$ denotes normal ordering with respect to the vacuum state. $A$ is the area of the 2DEG and $\kappa$ is the Thomas-Fermi cut-off parameter. 
Expressing \cref{eq:H_toe} in units of effective Bohr radius $a_B=\hbar^2/(m e^2)$ and Rydberg energy $\mathrm{Ry}=e^2/(2 a_B)$ allows to parameterize the 2DEG by the dimensionless Wigner-Seitz radius $r_s\,a_B=(\pi\,n)^{-1/2}$. The latter can be interpreted as average distance between electrons given the electronic density $n$. For high densities (${r_s\to 0}$), the kinetic term dominates, while at low densities (${r_s\to \infty}$) Coulomb interactions will take over. 
In the Hartree-Fock approximation, the groundstate is a Slater determinant state $\ket{\textrm{HF}}$ of plane waves and the quasiparticle has the energy dispersion 
\begin{equation}
\epsilon_{\vec{k}}^{\textrm{HF}}=\epsilon_{\vec{k}}+\Sigma^F_{\vec{k}},
\end{equation}
where $\Sigma^F_{\vec{k}}=-\sum_{|\vec{k'}|<k_F}V_{\vec{k}-\vec{k'}}$ is the exchange self-energy.

\begin{figure}
\centering
\includegraphics[width=1.0\linewidth]{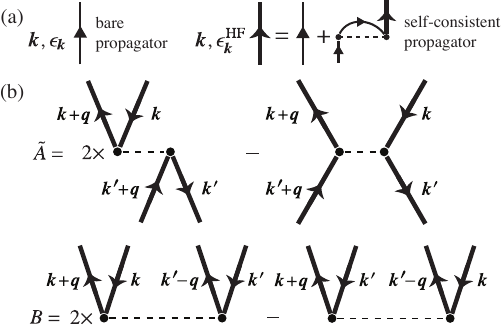}
\caption{
Diagrammatic interpretation for the matrix elements of $A$, $B$ in the excitation eigenequation for the 2DEG [c.f.~\cref{eq:RPA_eig}].  
(a)~Bare non-interacting propagator and self-consistent propagator accounting for Fock self-energy renormalization. 
(b)~TDHF matrix elements with direct and exchange contributions to the scattering ($\tilde{A}$) and to the double-excitations ($B$). 
Note that for $A$, we only show the contributions $\tilde{A}$ without the trivial kinetic term. Spin summation leads to factors of $2$ in the direct diagrams. 
}
\label{fig_diagrams}
\end{figure}

\begin{figure*}
\centering
\includegraphics[width=1.0\linewidth]{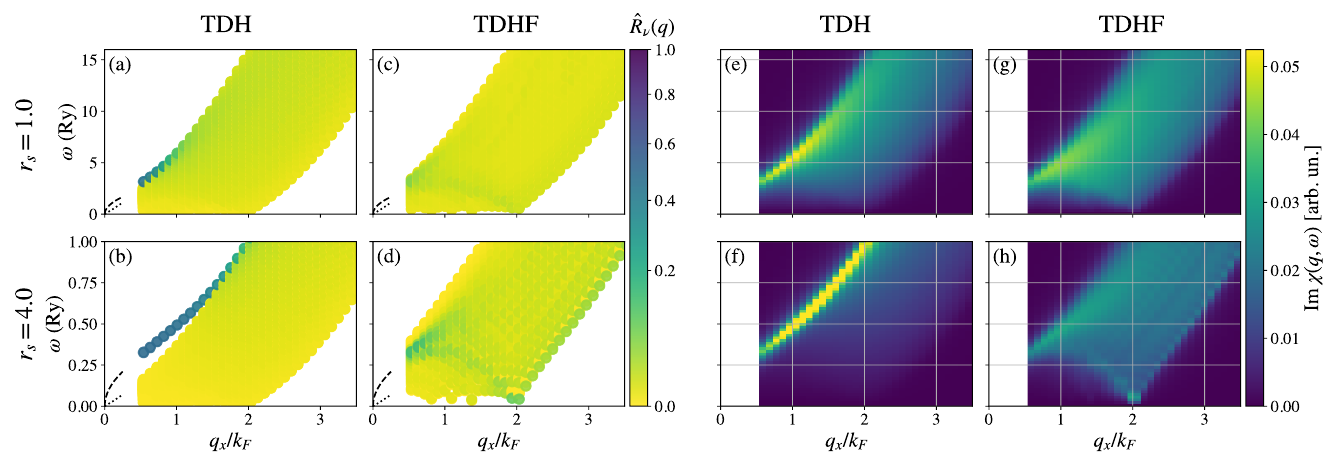}
\caption{
Spectrum, plasmon disperion and many-body stability of the 2DEG. (a--d) Excitation spectra, over an extended range of $ q $ including $ 2k_F $, obtained using the gRPA equations [cf.~\cref{eq:RPA_eig}] for time-dependent Hartree (TDH) and time-dependent Hartree-Fock with exchange self-energy (TDHF). (e--h) The corresponding charge susceptibilities for each scheme. Results are presented for electron gas density parameters $r_s = 1.0,4.0$. Notably, in the expanded $q$-range, TDHF shows a tendency to forming charge density waves, see main text.  Here, we used $dk=0.08 k_F$, $\kappa=0.1 dk$ and $q\geq 5 dk$.
}
\label{fig2}
\end{figure*}

\subsection{Generalized RPA (gRPA) and time-dependent mean-field}

Given the $N$-particle many-body groundstate $|\Psi\rangle$, a charge-neutral density excitation of momentum $\vec{q}$ can be written as $Q^\dagger_{\nu}(\vec{q}) |\Psi\rangle$, where 
\begin{align} \label{eq:exc_Qoperator}
    Q^\dagger_{\nu}(\vec{q})
    =
    \sum_{\vec{k}\in K_{\vec{q}}^+} 
    X_{\vec{k}\vec{q}}^{\nu} c^{\dagger}_{\vec{k}+\vec{q}}c^{\pdagger}_{\vec{k}} - 
    \sum_{\vec{k}\in K_{\vec{q}}^-}
    Y_{\vec{k}\vec{q}}^{\nu} c^{\dagger}_{\vec{k}} c^{\pdagger}_{\vec{k}-\vec{q}}.
\end{align}
Note that the spin index is being summed over and hence omitted~\footnote{This simply leads to a factor of $2$ in front of $V_q$ in \cref{eq:AB}}.
The summations account for all particle-hole pairs consistent with the Pauli-exclusion principle, described by ${K_{\vec{q}}^{\pm}=\{\vec{k} \,| \, \epsilon_{\vec{k}}\!<\! \epsilon_F \!<\! \epsilon_{\vec{k}\pm\vec{q}}\}}$. The number of pairs, $M_{\vec{q}}=|K_{\vec{q}}^{\pm}|$, is finite due to discretization. 
The first term in \cref{eq:exc_Qoperator} creates a particle-hole pair with momentum $+\vec{q}$ in $|\Psi\rangle$: $c^{\dagger}_{\vec{k}+\vec{q}}c^{\pdagger}_{\vec{k}}|\Psi\rangle$ where ${|\vec{k}|\!<\!k_F}$, ${|\vec{k}+\vec{q}|\!>\!k_F}$. The second term annihilates a particle-hole pair with momentum $-\vec{q}$ in $|\Psi\rangle$: $c^{\dagger}_{\vec{k}} c^{\pdagger}_{\vec{k}-\vec{q}}|\Psi\rangle$ where ${|\vec{k}|\!<\!k_F}$, ${|\vec{k}-\vec{q}|\!>\!k_F}$. 
Here and in what follows, holes are labeled with momentum $\vec{k},\vec{k}'$ and particles are labeled with ${\vec{k}\pm \vec{q}}$ and ${\vec{k}'\pm \vec{q}}$. 
Note that the Hartree-Fock state $|\text{HF}\rangle$ is the vacuum of the annihilation operator $Q_{\nu}(\vec{q})|\text{HF}\rangle=0$ only when we set $Y=0$. The many-body state annulled by $Q_{\nu}(\vec{q})$ when $Y\neq0$ is called generalized RPA (gRPA) state, i.e.,
\begin{equation} \label{eq:RPA_groundstate}
    Q_{\nu}(\vec{q}) |\text{gRPA}\rangle=0.
\end{equation}
In contrast to state $|\text{HF}\rangle$, which is void of any particle-hole pairs, state $|\text{gRPA}\rangle$ must (by definition) inherently incorporate such pairs, resulting in a better approximation to the ground state. 
The energy required to create a quantum of excitation, denoted as $\hbar\omega_\nu(\vec{q})$, can be determined  by applying the equation of motion 
\cite{peter1980nuclear,schuck2021equation} to the state
$Q_\nu^{\dagger}(\vec{q})|\text{gRPA}\rangle$. This leads to the following equation with double commutators:
\begin{widetext} 
\begin{align} \label{eq:eom}
   \langle\text{gRPA}|[c^\dagger_{\vec{k}'}c^{\pdagger}_{\vec{k}'+\vec{q}},[H, Q^\dagger_{\nu}(\vec{q})]]|\text{gRPA}\rangle &=\hbar \omega_\nu(\vec{q}) 
   \langle\text{gRPA}|[c^\dagger_{\vec{k}'}c^{\pdagger}_{\vec{k}'+\vec{q}}, Q^\dagger_{\nu}(\vec{q})]|\text{gRPA}\rangle \nonumber \\
    \langle\text{gRPA}|[c^\dagger_{\vec{k}'-\vec{q}}c^{\pdagger}_{\vec{k}'},[H, Q^\dagger_{\nu}(\vec{q})]]|\text{gRPA}\rangle &=\hbar \omega_\nu (\vec{q}) \langle\text{gRPA}|[c^\dagger_{\vec{k}'}c^{\pdagger}_{\vec{k}'+\vec{q}}, Q^\dagger_{\nu}(\vec{q})]|\text{gRPA}\rangle.  
\end{align}
Next, we make an approximation to replace $|\textrm{gRPA}\rangle$ with $|\textrm{HF}\rangle$ in \cref{eq:eom}. This substitution leads to an eigenvalue equation for the particle-hole pairs, known as the gRPA eigenvalue equation \footnote{The gPRA equation can also be derived in time-dependent mean-field theory \cite{peter1980nuclear}.},
\begin{align} \label{eq:RPA_eig}
\sum_{\vec{k}}
\mathcal{M}_{\vec{k'},\vec{k}}(\vec{q})
\begin{bmatrix}
X^{\nu}_{\vec{kq}} \\
Y^{\nu}_{\vec{kq}}
\end{bmatrix} 
= \sum_{\vec{k}}
\begin{bmatrix}
 A_{\vec{k}'+\vec{q},\vec{k'};\,\vec{k}+\vec{q},\vec{k}}   & B_{\vec{k}'+\vec{q},\vec{k'};\,\vec{k}-\vec{q},\vec{k}}   \\
(B_{\vec{k}'-\vec{q},\vec{k'};\,\vec{k}+\vec{q},\vec{k}})^*   & (A_{\vec{k}'-\vec{q},\vec{k'};\, \vec{k}-\vec{q},\vec{k}})^* 
\end{bmatrix}
\begin{bmatrix}
X^{\nu}_{\vec{kq}} \\
Y^{\nu}_{\vec{kq}} 
\end{bmatrix} 
= \hbar \omega_\nu(\vec{q})
\begin{bmatrix}
1   & 0  \\
0  & -1
\end{bmatrix}
\begin{bmatrix}
X^{\nu}_{\vec{k'q}} \\
Y^{\nu}_{\vec{k'q}}
\end{bmatrix},
\end{align}
\end{widetext}
where the matrix elements of the matrix $\mathcal{M}$ are given by
\begin{align} \label{eq:AB} 
     A_{\vec{k}'+\vec{q},\vec{k'};\vec{k}+\vec{q},\vec{k}} 
     &= (\epsilon_{\vec{k}+\vec{q}}+
     \Sigma^F_{\vec{k}+\vec{q}}
     -\epsilon_{\vec{k}}
     -\Sigma^F_{\vec{k}})\delta_{\vec{k},\vec{k}'} + 2 V_{\vec{q}} - V_{\vec{k}-\vec{k}'}, \nonumber \\
     B_{\vec{k}'+\vec{q},\vec{k'};\vec{k}-\vec{q},\vec{k}} 
     &=  2 V_{\vec{q}} - V_{\vec{k}-(\vec{k}'-\vec{q})}.
\end{align}
It turns out that ${|\textrm{gRPA}\rangle\mapsto|\textrm{HF}\rangle}$ in \cref{eq:eom} implies the quasi-boson approximation in \cref{eq:QBA}, which we will make use of later in \cref{sec:corr_energy}.

For a specific $\vec{q}$, \cref{eq:RPA_eig} represents a $2M_{\vec{q}}\times2M_{\vec{q}}$ eigenvalue problem, where $M_{\vec{q}}$ denotes the number of points in the set $K_{\vec{q}}^{+}$ (or $K_{\vec{q}}^{-}$). 
The matrix elements of $\mathcal{M}(\vec{q})$ represent various scattering amplitudes for the mean-field quasiparticles, shown as  Feynman diagrams in \cref{fig_diagrams}. 
The matrix $\mathcal{M}(\vec{q})$ captures both the direct and the exchange scattering between particle-hole pairs with momentum $+{\vec{q}}$ and $-\vec{q}$. 
Matrix $A$ characterizes the scattering between two pairs in $K_{\vec{q}}^{+}$ or two pairs in $K_{\vec{q}}^{-}$. Conversely, matrix $B$ describes the scattering involving one particle-hole pairs from $K_{\vec{q}}^{+}$ and another from $K_{\vec{q}}^{-}$. The simplicity and similarity of matrix elements in $A$ and $B$ results from to the fact that the Hartree-Fock quasiparticles in the 2DEG are just plane wave states. In systems with additional orbital degrees of freedom, form factors would enter the matrix elements.

\begin{figure*}
\centering
\includegraphics[width=1.0\linewidth]{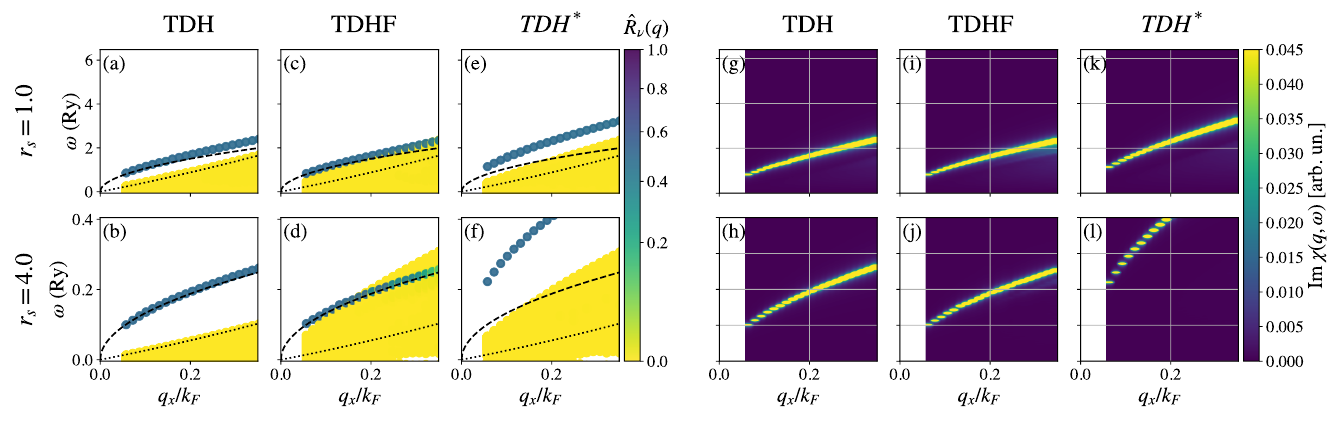}
\caption{
Long-wavelength spectrum and plasmon modes of the 2DEG. (a--f) Excitation spectra obtained as eigensolutions of the gRPA equations [cf.~\cref{eq:RPA_eig}] for three different approximation schemes: time-dependent Hartree (TDH), time-dependent Hartree-Fock with exchange self-energy (TDHF), and time-dependent Hartree with exchange self-energy (TDH$^*$). (g--l) The corresponding charge susceptibilities for each scheme. All results are plotted for two electron gas density parameters, $r_s=1.0,4.0$, as indicated to the left of each row. Color intensities in the spectra and susceptibilities represent the magnitude of the spectral weight $\hat{R}_{\nu}(q)$ and the imaginary part of the charge susceptibility \(\chi(q, \omega)\), respectively. The plasmon modes for TDH and TDHF are comparable as a result of a conserving approximation. Note that TDH$^*$ gives unphysical results, see main text. 
Here, we used $dk=0.006 k_F$, $\kappa=0.15 dk$ and $q\geq 3 dk$.
}
\label{fig3}
\end{figure*}

The entirety of the eigenspectrum of \cref{eq:RPA_eig} provides the basis for constructing the spectral representation of the particle-hole Green function \cite{thouless2014quantum}. This foundation allows us to express the density-density response function as
\begin{equation}
\chi(\vec{q},\omega) =
\sum_{\nu=-M_{\vec{q}}}^{M_{\vec{q}}} \text{sgn}(\omega_\nu(\vec{q}))\frac{R_{\nu}(\vec{q})}{\omega-\omega_\nu(\vec{q})+i0^+}.
\end{equation}
Here, $R_{\nu}(\vec{q})$ represents the modulus squared of the sum of all entries of an gRPA eigenvector, defined as
\begin{equation}
R_{\nu}(\vec{q}) \equiv  |\langle \nu \vec{q}|\rho_{\vec{q}}|0\rangle|^2 = \bigg| \sum_{\vec{k}=1}^{M_{\vec{q}}}(X^\nu_{\vec{kq}}+Y^\nu_{\vec{kq}}) \bigg|^2,
\end{equation}
which is equivalent to the probability of finding one quanta of density excitation in the excited state denoted by $|\nu\vec{q}\rangle$. In this notation, $R_{\nu}(\vec{q})$ is proportional to the residue of $\chi(q,\omega)$ at its simple pole $\omega_{\nu}(q)$, serving as a metric for measuring the collectivity of a particular gRPA eigenvector. Particularly, in cases where the excitation is an equal superposition of all particle-hole pairs, such as when $X_{\vec{kq}}^{\nu}+Y_{\vec{kq}}^{\nu}=1/M_{\vec{q}}$, $R_{\nu}(\vec{q})$ peaks at unity. Thus, $R_{\nu}(\vec{q})$ allows to distinguish between collective excitations and uncorrelated particle-hole pair excitations, especially when the former merges into the particle-hole continuum.

\subsection{Results: Charge-neutral excitation spectrum and density-density response} \label{sec:results}

In this subsection, we designate the eigenvalue spectrum derived from Eq.~\eqref{eq:RPA_eig} as the time-dependent Hartree-Fock (TDHF) approximation and proceed to compare it with two other prevalent approximations. 
The first of these is the time-dependent Hartree (TDH) approximation, which involves setting both the exchange self-energy and exchange-scattering to zero,i.e.~$\Sigma_k^F=V_{\vec{k},\vec{k'}}=
V_{\vec{k},\vec{k'}-\vec{q}}=0$ in \cref{eq:AB}. This approach is equivalent to the RPA approximation introduced by Bohm and Pines. The second approach, denoted TDH$^{*}$ approximation here, differs slightly; here, we retain the exchange self-energy while setting the exchange scattering to zero, i.e.~ $\Sigma_k^F\neq0$, 
$V_{\vec{k},\vec{k'}}=V_{\vec{k},\vec{k'}-\vec{q}}=0$ in \cref{eq:AB}. Note both the TDHF and TDH are classified as conserving approximations, following Baym's criteria \cite{baym1962self}, whereas the TDH$^{*}$ is not.

Figs.~\ref{fig2}(a) and (b) present the excitation spectrum computed using the TDH approximation for $r_s=1$ and $r_s=4$, respectively. This spectrum consists of a continuum of particle-hole excitations, which exhibit small $R_{\nu}(q)$ values, and one plasmon mode characterized by significantly large $R_{\nu}(q)$ values. As $r_s$ increases, the plasmon mode progressively merges into the continuum at larger $q$ values.

Figs.~\ref{fig2}(c) and (d) depict the excitation spectrum determined within the TDHF approximation for $r_s=1$ and $r_s=4$. Notably, the plasmon mode is fully merged into the particle-hole continuum. Moreover, the near-zero-frequency excitations near $2k_F$ exhibit an enhanced $R_{\nu}(q)$. 
A deeper insight into these effects can be obtained by analyzing the spectral weight of the density fluctuations in Fig.~\ref{fig2}(h).

In the TDH approximation, the spectral weight is concentrated around the plasmon mode. This concentration becomes more pronounced with increasing $r_s$ due to the larger energy separation between the particle-hole continuum and the plasmon mode. Conversely, in the TDHF approximation, the spectral weight is more uniformly spread across the continuum. A notable peak in spectral weight is observed at wavenumber $q=2k_F$. This mode represents an inherent instability of the electron gas towards the formation of a charge (or spin) density wave \cite{overhauser1962spin,overhauser1978charge}.
As $r_s$ increases, this mode gains more spectral weight and moves closer to zero-energy. When this mode attains zero energy, we have to consider the charge-density wave as competing groundstate, and subsequently, examine the excitation spectrum based on this state. The inclusion of screening in our long-range Coulomb interaction, controlled by $\kappa\simeq0.1 dk$ ($dk$ is the momentum space discretization), suppresses the charge density wave instability. For a discussion on how the screening constant curbs the density wave instability in a three-dimensional electron gas, see Ref.~\cite{kranz1984exchange}.

In Fig.~\ref{fig3}, we focus on the small $q$ region of the particle-hole excitation spectrum. We use broken lines to indicate the classical plasmon-dispersion 
\begin{equation} \label{eq:plasmon_dispersion}
    \omega_{pl}(q) = \frac{\sqrt{2q}}{r_s},
\end{equation}
and the maximum of the particle-hole continuum computed from the parabolic dispersion
\begin{equation} \label{eq:ph_dispersion}
    \text{max}(\epsilon_{\vec{k}+\vec{q}}-\epsilon_{\vec{k}}) = 2k_Fq+q^2.
\end{equation}
Figs.~\ref{fig3}(a) and (b) demonstrate that the TDH approximation closely aligns with \cref{eq:plasmon_dispersion,eq:ph_dispersion}. There is an extensive $q$-range where the plasmon mode is spectrally isolated from the uncorrelated particle-hole spectrum. 
Figs.~\ref{fig3}(c) and (d) show the spectrum obtained in the TDHF approximation. While the plasmon dispersion remains aptly described by \cref{eq:plasmon_dispersion}, the boundary of the particle-hole continuum shifts notably up compared to \cref{eq:ph_dispersion}. 
The upward shift of the particle-hole continuum boundary arises because the exchange self-energy of the holes (i.e., occupied states) are more negative than that of the particle state (i.e., unoccupied state). Consequently, within the specified $q$-range in \cref{fig3}, the plasmon mode undergoes weak Landau damping. 
To provide a contrasting perspective, Figs.~\ref{fig3}(e) and (f) show the excitation spectra for TDH$^{*}$. As discussed above, TDH$^{*}$ is not a conserving approximation: the exchange effect is only retained in the single-particle propagator but omitted in the vertices \cite{baym1962self}. While it describes the particle-hole continuum effectively, it inadequately captures the plasmon mode dispersion. 

Figs.~\ref{fig3}(g--l) show the spectral function computed with different approximation schemes. The most important feature in these plots is that the spectral weight is almost completely exhausted by by the plasmon dispersion at small $q$. This characteristic persists even when the plasmon dispersion slightly overlaps with the continuum, as shown in Fig.~\ref{fig3}(d). This observation signals the validity for a powerful approximation that effectively describes the long-wavelength limit of the dynamical structure factor in the electron liquid. 
The method is known as ``single-mode approximation'' such that  $\text{Im}\chi(q,\omega)/\pi \sim f(q)\delta(\omega-\omega_{pl}(q))$.

\section{Quasi-boson approximation and correlation energy}
\label{sec:corr_energy}

In this section, we use the gRPA eigenvalue spectrum in \cref{sec:method} to compute the correlation energy using a quasi-boson approximation. 
To begin, let us revisit some fundamental properties of the bosonic quadratic Hamiltonian
\begin{equation} \label{eq:K}
    K=A b^\dagger b+ \frac{1}{2} (B b^\dagger b^\dagger +  B^* bb), 
\end{equation}
where $A$ and $B$ are parameters, and $b^{\dagger}$ and $b$ are bosonic creation and annihilation operators.
Due to its quadratic nature, the double commutator of $K$ with the creation and annihilation operators yields numerical values instead of operators.
To transform $K$ into the diagonal form $K=\epsilon \alpha^\dagger \alpha$, where $[\alpha,\alpha^\dagger]=1$, we define $\alpha^\dagger=Xb^\dagger-Yb$ and evaluate its double commutator with $K$,
\begin{align}  \label{eq:commutator}
    [\alpha,[K,\alpha^\dagger]]=&   
[X^* \,,\,Y^*]
\begin{bmatrix}
[b,[K,b^{\dagger}]]   & -[b,[K,b]]  \\
-[b^\dagger,[K,b^{\dagger}]]  & [b^\dagger,[K,b]]
\end{bmatrix} 
\begin{bmatrix}
X \\
Y
\end{bmatrix} \nonumber \\
=&[X^* \,,\,Y^*]
\begin{bmatrix}
A   & B  \\
B^*  & A
\end{bmatrix} 
\begin{bmatrix}
X \\
Y
\end{bmatrix}=\epsilon.
\end{align}
Using $|X|^2-|Y|^2=1$, \cref{eq:commutator} can be written as the following eigenvalue equation
\begin{equation} \label{eq:K_eig}
    \begin{bmatrix}
A   & B  \\
B^*  & A
\end{bmatrix} 
\begin{bmatrix}
X \\
Y
\end{bmatrix}=\epsilon \begin{bmatrix}
X \\
-Y
\end{bmatrix},
\end{equation}
which is structurally identical to the gRPA equation in \cref{eq:RPA_eig}.

Next, we implement the quasi-boson approximation by assuming the fermion bilinears $c^\dagger_{\vec{k}}c^{\pdagger}_{\vec{k}+\vec{q}}=b_{\vec{k}+\vec{q},\vec{k}}$ satisfy the bosonic commutation relation 
\begin{equation}
    [b_{\vec{k}+\vec{q},\vec{k}}, b^{\dagger}_{\vec{k'}+\vec{q},\vec{k'}}]=\delta_{\vec{k},\vec{k}'}.
\end{equation}
A bosonic Hamiltonian, denoted as $H_B$, is then constructed by equating the double commutators of $H_B$ with the quasi-boson creation and annihilation operators to the matrix elements of the gRPA equation, i.e., 
\begin{subequations}
\begin{align}  
    [b_{\vec{k}'+\vec{q},\vec{k}'},[H_B, b^{\dagger}_{\vec{k}+\vec{q},\vec{k}}]]
     &= A_{\vec{k}'+\vec{q},\vec{k'};\vec{k}-\vec{q},\vec{k}}, \\
    [b_{\vec{k}'+\vec{q},\vec{k}'},[H_B, b_{\vec{k}+\vec{q},\vec{k}}]]
     &=  -B_{\vec{k}'+\vec{q},\vec{k'};\vec{k}-\vec{q},\vec{k}}.
\end{align}
\end{subequations}
Analogous to the example in the beginning of this section, see \cref{eq:K,eq:commutator,eq:K_eig}, we find
%
\begin{align}\label{eq:Hamiltonian_quasi-boson}
    H_{B} =&
    \sum_{\vec{q},\vec{k},\vec{k}'}\,\bigg(
A_{\vec{k}'+\vec{q},\vec{k'};\vec{k}+\vec{q},\vec{k}} \;
    b^{\dagger}_{\vec{k'}+\vec{q},\vec{k'}} b_{\vec{k}+\vec{q},\vec{k}} 
    \nonumber \\
    &\quad 
    +\frac{1}{2}B_{\vec{k}'+\vec{q},\vec{k'};\vec{k}-\vec{q},\vec{k}}  \;
    b_{\vec{k'}-\vec{q},\vec{k'}}
    b_{\vec{k}+\vec{q},\vec{k}} + \text{c.c.}\bigg) 
    \\
    =&\frac{1}{2}\sum_{\vec{k},\vec{k'},\vec{q}}
    \begin{bmatrix}
    b^{\dagger}_{\vec{k'}+\vec{q},\vec{k'}} &
    b_{\vec{k'}-\vec{q},\vec{k}'}
    \end{bmatrix}
    \mathcal{M}_{\vec{k'},\vec{k}}(\vec{q})
    \begin{bmatrix}
    b_{\vec{k}+\vec{q},\vec{k}} \\
    b^{\dagger}_{\vec{k}-\vec{q},\vec{k}}
    \end{bmatrix}
    %
    - \frac{1}{2} \mathrm{tr}\, A. \nonumber 
\end{align}
%
In the second line, we used the bosonic commutation relation to express $b^\dagger b $ as $ \frac{1}{2}(b^\dagger b+b b^\dagger -1)$. Here, $\mathcal{M}_{\vec{k'},\vec{k}}(\vec{q})$ is defined in Eq.~\eqref{eq:RPA_eig}. Leveraging the eigenvalue spectrum of $\mathcal{M}_{\vec{k'},\vec{k}}(\vec{q})$, the bosonic Hamiltonian takes the form
\begin{align}
    H_B&= \frac{1}{2}\sum_{\vec{q}}\sum_{\nu=1}^M \hbar \omega_{\nu}(\vec{q})[Q_{\nu}^\dagger (\vec{q})Q_{\nu}(\vec{q})+ Q_{\nu}(\vec{q})Q_{\nu}^\dagger(\vec{q}) ] -\! \frac{\mathrm{tr}\, A}{2}  \nonumber \\
    &=
     \frac{1}{2}\sum_{\vec{q}}\sum_{\nu=1}^M  \hbar \omega_{\nu}(\vec{q})Q_{\nu}^\dagger (\vec{q})Q_{\nu}(\vec{q}) + E_{\text{zp}},
\end{align}
where $Q^\dagger_{\nu}(\vec{q})$ is defined in \cref{eq:exc_Qoperator}. Here $\nu=1,\dots,M$ labels the positive eigenvalues. The Hamiltonian $H_B$ describes a portion of the residual interaction within mean-field theory as a collection of harmonic oscillators and the zero-point energy $E_{\text{zp}}$. The groundstate energy of $H_B$ is 
\begin{equation}
    E_{\textrm{zp}}=\frac{1}{2} \sum_{\vec{q}}\sum_{\nu=1}^{M} \hbar \omega_{\nu}(\vec{q}) - \frac{1}{2}\sum_{\vec{k},\vec{q}} A_{\vec{k}+\vec{q},\vec{k};\vec{k}+\vec{q},\vec{k}}.
\end{equation}
In \cref{fig1}, we show the total correlation energy per particle $
E_{\textrm{corr}} =E_{\textrm{zp}}/N$ at different $r_s$ where $N=n_eA$ is the number of electrons. It is noteworthy that our results show good agreement with those obtained from QMC studies as well as the STLS approximation.
%

\section{Discussion and outlook}
\label{sec:conclusion}

To summarize, we have studied the charge-neutral particle-hole excitation spectrum and response function of the 2DEG by formulating  the generalized RPA eigenvalue equation for extended systems. We used the resulting excitation spectrum to evaluate the 2DEG correlation energy by invoking the quasi-boson approximation. The gRPA and QBA method presented in this work has a convenient diagrammatic construction and systematically accounts for both direct scattering and exchange scattering (i.e., TDH and TDHF). 

In stark contrast, the most commonly-used RPA method based on coupling constant integration is exclusively a TDH method \cite{giuliani2005quantum}, i.e., it does not account for exchange effects. In that method, the susceptibility simplifies to $\chi_{\lambda}(\vec{q},\omega) = \chi_0(\vec{q},\omega)/(1-\lambda\, V_{\vec{q}}\chi_0(\vec{q},\omega))$, where $\chi_0$ is the non-interacting susceptibility, $\lambda$ the coupling constant. The correlation energy $E_{\text{corr}}$ is given analytically through coupling-constant integration involving $\chi_{\lambda}(\vec{q},i\omega)$. While such a momentum-local RPA approximation is powerful for analytical studies, it (1) is only strictly valid at high densities ($r_s\ll1$) and (2) can only be extended to include exchange effects by introducing semiclassical self-consistent local field factors. Such local-field methods are not easily extended to multiorbital systems and their validity is not easily checked. Our gRPA and QBA method has none of these limitations. However, our method requires careful diagrammatic construction, choice of discretization, particle-hole basis truncation, and large-scale numerical diagonalization. As we showed in this work, these issues can be overcome, leading to results that rival STLS and QMC methods.


Recent discoveries of correlated states and superconductivity in multilayer graphene \cite{zhou_half_2021,zhou_isospin_2021,zhou_superconductivity_2021} and Wigner crystals in TMDs \cite{morales2021metal,xu2020correlated,huang2021correlated,regan2020mott,matty2022melting,zhou2021bilayer} provide strong motivation to study gRPA correlation effects in these materials. In particular, this may allow to understand why their mean-field descriptions tend to be have some systematic issues. In this context, we highlight that the formalism in this work straightforwardly generalizes to multicomponent 2DEG-like problems, such as spin, valley and sublattice in the continuum models of graphene and TMD multilayers. It also generalizes to other types of particle-hole excitation spectra and response functions, including those for spin- and valley-flip excitations. 

To conclude, this work opens exciting new possibilities by providing a practical way to explore and understand correlations in complex two-dimensional condensed matter systems.

\begin{acknowledgments}
T.M.R.W.~is grateful for the financial support from the Swiss National Science Foundation (Postdoc.Mobility Grant No.~203152) and for the hospitality of the University of Kentucky, where part of this work was done.
\end{acknowledgments}

\bibliography{references}

\end{document}